\documentclass{article}[10pt]
\usepackage{graphicx}
\begin{document}
\title{A Note on Infinities in Eternal Inflation}
\author{George F. R. Ellis \and William R. Stoeger}
\maketitle
\begin{abstract}
In some well-known scenarios of open-universe eternal inflation,
developed by Vilenkin and co-workers, a large number of universes
nucleate and thermalize within the eternally inflating
mega-universe. According to the proposal, each universe nucleates
at a point, and therefore the boundary of the nucleated universe
is a space-like surface nearly coincident with the future light
cone emanating from the point of nucleation, all points of which
have the same proper-time. This leads the authors to conclude that
at the proper-time $t = t_{nuc}$ at which any such nucleation
occurs, an infinite open universe comes into existence. We point
out that this is due entirely to the supposition of the nucleation
occurring at a single point, which in light of quantum cosmology
seems difficult to support. Even an infinitesimal space-like
length at the moment of nucleation gives a rather different result
-- the boundary of the nucleating universe evolves in proper time
and becomes infinite only in an infinite time. The alleged
infinity is never attained at any finite time.
\end{abstract}

\section{Introduction}

It is commonly stated that infinities can occur in inflationary
cosmology in two ways. Firstly, it is stated that quantum tunneling
can lead from a universe with finite spatial sections to an imbedded
expanding universe domain with infinite spatial sections. Secondly,
it is claimed this will occur an infinite number of times: ``most
important of all is that once inflation starts it produces not just
one universe, but an infinite number of universes'' \cite{guth07}.
Thus what is proposed is the existence of an infinite number of
expanding universe domains like the one we see around us, but each
with somewhat different characteristics, all imbedded in a
surrounding inflationary universe; and each with infinite spatial
sections.\\

The way this is stated implies, implicitly or explicitly, that the
infinite spatial sections come into being instantaneously in each
bubble; as soon as a bubble nucleates in an inflating
universe, negatively curved infinite spatial sections come into
existence at that instant. This claim, which is repeated and
provides the basis for further philosophical speculations in later
work (e. g., Vilenkin and Garriga \cite{garvil01}, Knobe, Ohm, and
Vilenkin \cite{knoetal03}, and Vilenkin \cite{vil06}), is based
essentially on identifying the future light cone of the point of
nucleation - or more precisely, an exactly constant proper time
surface very close to that light cone - as the effective boundary
of the bubble universe. As such it is a surface of infinite extent
that is everywhere at the same spacetime distance from the origin,
with a foliation of spacelike surfaces of constant negative
curvature and infinite extent asymptotic to this light cone in its
future. However, this conclusion is the result of assuming that
the bubble universe originates at a single point, instead of as a
region with some extension, no matter how small. When we
substitute this alternative, the instantaneous emergence of the
infinite bubble boundary is no longer apparent. We are left with a
finite boundary which expands forever -- but only becomes infinite
in an infinite time, and hence is never in fact attained.\\

Thus in both cases, there is an ongoing process: at any finite
time there never exists the claimed infinite number; it is what
the situation tends to but never attains. The universe is evolving
towards such a state, but never reaches it; for that is the
essential nature of infinity.

\section{Origination of Infinite Bubble Universes: The Null Cone}

In their development of open-universe inflation, A. Vilenkin and
S. Winitzki \cite{vilwin97} consider a mega-universe dominated by
false vacuum inflating with an approximately de Sitter metric
$$ds^2 = -dt^2 + \exp( 2H_0 t)(dr^2 + r^2 d\Omega^2), \eqno (1) $$
where $H_0 = \sqrt{2\pi V_0/3}$, $V_0$ being the false vacuum
potential. Within this inflating mega-verse, spherical bubbles of
true vacuum nucleate to form daughter universes which quickly
thermalize and evolve separately as
Friedmann-Lema\^{i}tre-Robertson-Walker (FLRW) universes. If we
focus on one case of cosmic nucleation,  at the moment of
nucleation -- at a particular point in time and in space, say at
$t = 0$ and $r = r_0$ -- a bubble of true vacuum is formed, which
looks like an open FLRW universe with
$$ds^2 = d\tau^2 + a^2(\tau)(d\xi^2 + sinh^2 \xi d\Omega^2), \eqno (2)$$
where for small values of $\tau$, setting nucleation at $\tau = t
= 0$, the bubble has yet to thermalize and is still inflating. So
it is very close to de Sitter, with
$$a(\tau) = H_0^{-1}sinh (H_0\tau). \eqno(3)$$
Vilenkin and Winitzki \cite{vilwin97} carefully  relate the
coordinates of the metrics, discuss the conditions for continued
brief acceleration of the bubble universe and for its
thermalization. Similar analyses are given by other writers, for
example Freivogel et al \cite{frei07}, but in that case with
tunneling from the compact $k = +1$ form of the de Sitter universe. \\

The argument for an infinite bubble universe originating at a
point $t=\tau=0$ with point-like radial position $r = r_0$ is
straightforward. Since it has no spatial extension, from equation
(1) $ds^2 = -dt^2$, and the proper interval $s$ there can be
identified with $t = 0$. So $s = t = 0$ at the origin. But the
whole future light cone emanating from there is at the same proper
distance from the origin, since it is given by $ds = 0$. And that
surface extends to infinity. In a definite sense then, that whole
surface comes into existence instantaneously at the instant $t =
0$. Vilenkin and Winitzki \cite{vilwin97} show that the boundary
of the bubble universe is in this case a constant proper-time
spacelike surface infinitesimally close to that future light cone.
Thus, an infinite bubble universe comes into existence at that
moment. Spacelike surfaces of constant proper time occur within
this boundary, each at a constant proper time from the origin;
thus after say a finite time $\tau_1$, no matter how small,
infinite spacelike surfaces exist with every point produced
through physical processes taking the identical time $\tau_1$
since nucleation, no matter how far from the origin they are. An
actual infinite spatial extent comes into existence in a finite
time, originating from a universe with compact spatial sections -
a truly remarkable claim, based on the nature of the hyperbolic
geometry of special and general relativity theory.

\section{Spatially Extended Finite Alternative}

Is this claim that an infinite bubble universe emerges
instantaneously at $t=\tau = 0$ supportable? As emphasized above,
this rests on the supposition that the surface of the bubble
universe can be identified with the future light cone emanating
from that point, which requires that it originate precisely at
that point -- without spatial extension. However, this supposition
seems somewhat dubious from the point of view of quantum
cosmology, taking into account the fact that such transitions
really take place within finite spatial volumes -- not at
idealized points. Indeed according to Freivogel et al
\cite{frei07}, quantum tunnelling leads to a domain of finite
radius coming into existence instantaneously. This is quite
different than when it is supposed to occur at a point-like event.  \\

So, what happens when we suppose, instead, that the bubble
universe originates in a volume with some finite extension? Let's
simply represent it as $\Delta r^2$. Then, from the point of view
of the de Sitter mega-universe metric, we see that that
originating surface is not at the same proper distance from the
origin, since
$$\Delta s^2 = -\Delta t^2 + \exp (2H_0t)\Delta r^2.  $$
Clearly $t=\tau = 0$ cannot now be identified with the
proper-interval of the surface of the originating bubble, and, as
that surface evolves forward in time, expanding rapidly by
inflation at least for a short while before thermalization, its
boundary will not have the same proper distance from every part of
that originating surface at $t= \tau =0$. In fact, the proper time
on that boundary measured from the various spatial points in the
nucleated bubble, will diverge. There is then no support for
considering it instantaneously infinite -- it began finite, and
continues to be finite at every physically attainable time. It
approaches infinity in an infinite time -- but that is never
actually reached or completed,
in any real sense of that word. \\

Thus, the interpretation of an infinite $t= 0$ constant initial
surface of a bubble universe coming into existence at that time
within an eternally inflating mega-universe appears to be an
artifact of treating its origination in a unwarranted idealized
fashion -- at a single space-time point. This is an example of how
careful we need to be in drawing conclusions from idealizations
which involve unextended points and the infinities which are
connected with them.

\section{Spacelike Surfaces and the Emerging Universe}

The key underlying point is that the whole spacetime does not come
into being instantaneously: things take place physically, as events
unroll along particle world lines (particles tunnel, scalar fields
roll down potential surfaces, particles collide, etc). The outcome
is determined as it happens (because quantum events are involved),
determining both what happens in the space-time and (because
space-time curvature is determined by the matter in the space-time)
even the space-time structure itself. We do not in the real world
have a block universe that instantaneously comes into being; rather
we have an Evolving Block Universe that unrolls over time
\cite{ell06}.\\

 But which spacelike surfaces are the relevant ones for this process?
 Different choices of time are represented by different choices of
 spacelike surfaces of constant time, as depicted in Figure 1 for
 the case of Minkowski (flat) spacetime, which adequately illustrates
 the main points to be made here. Spacelike surfaces $S_0 = constant$
 are at constant proper time from the origin of coordinates. In
effect, the suggestion made by Vilenkin et. al. is that we should
take seriously these surfaces of constant distance from the origin
rather than the surfaces of constant Minkowski time $T$.\\

\begin{figure}
\includegraphics[width=15cm]{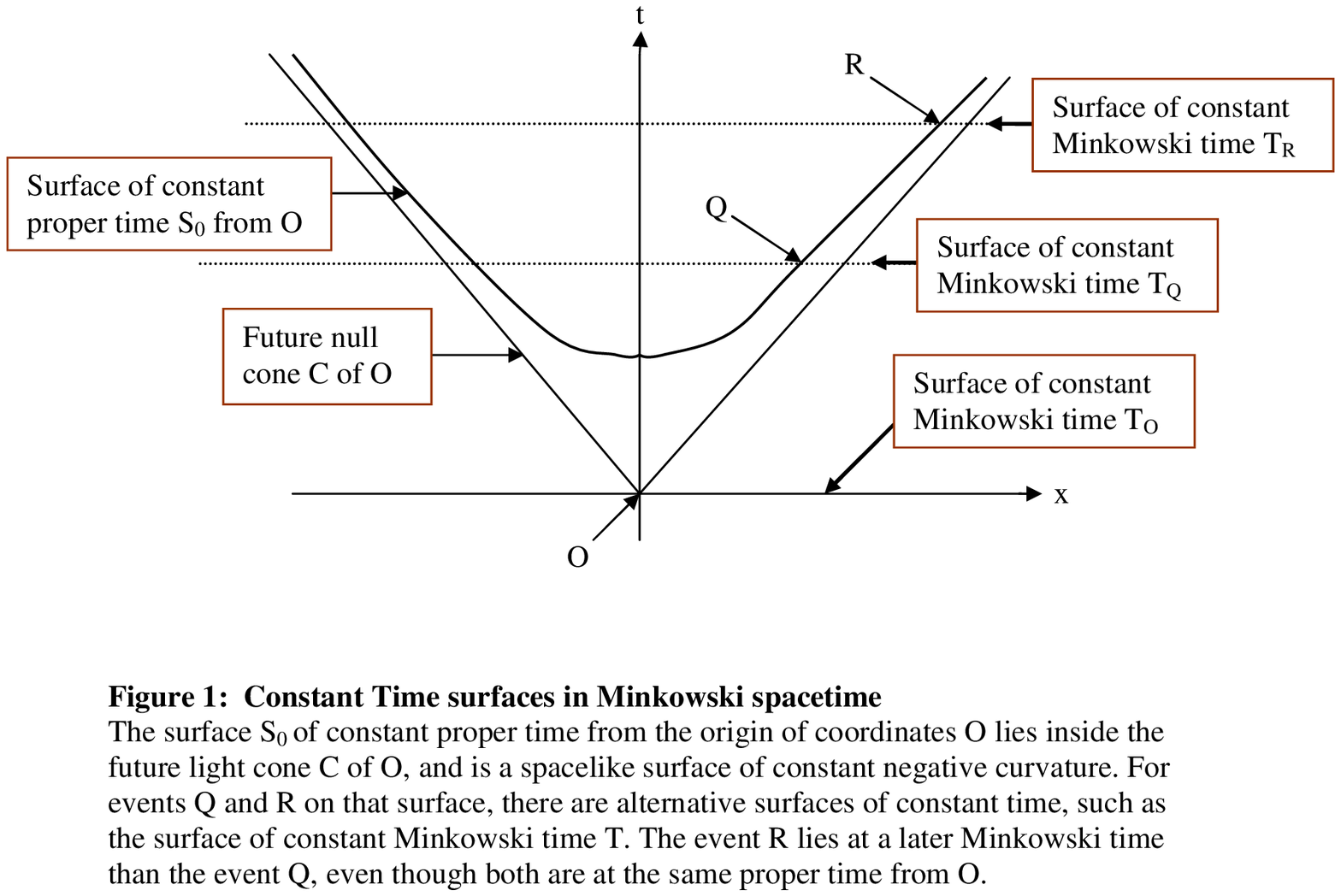}
\end{figure}

How to choose between them? Well, in fact \cite{ell06} events
normally unroll along timelike or null world lines, rather than
being based on spacelike surfaces, convenient as these are for
setting up coordinate systems. Every point in each surface $S_0 =
constant$ is at the same proper time from the origin of
coordinates. So one can imagine a physical process that takes a
proper time $S_0$ to develop after the initiation event at O; then
the outcome will occur simultaneously at every point on the
surface $S_0$= constant, no matter how far away in spatial terms.
Thus one can claim that infinitely far away events (in spacelike
terms) occur simultaneously with the nearest ones, in these
surfaces. This is the rationale for claiming infinities of spatial
events actually exist immediately after the tunnelling; for this
argument holds no matter how small $S_0$ is.\\

It is in this context that it matters that the causal origin of the
physical events after tunneling is of finite extent, see Figure 2 of
\cite{frei07} for this context. It is as a result of this finite
size that infinities never exist at any finite time after the
tunneling event. We consider this first in the paradigmatic case of
Minkowski spacetime, and then in the context of general bifurcate
Killing horizons, such as occur for example in de Sitter spacetime.

\subsection{Paradigmatic Example: Minkowski Spacetime}

Proper time from the origin O to a point Q on the surface at
constant proper time from O in Minkowski spacetime is by definition
of that surface the same as the proper time from the origin O to any
other point R on the surface (See Figure 2), even though the first
corresponds to a smaller Minkowski time T than second. However the
proper time from P to R is greater than the proper time from P to Q.
\begin{figure}
\includegraphics[width=15cm]{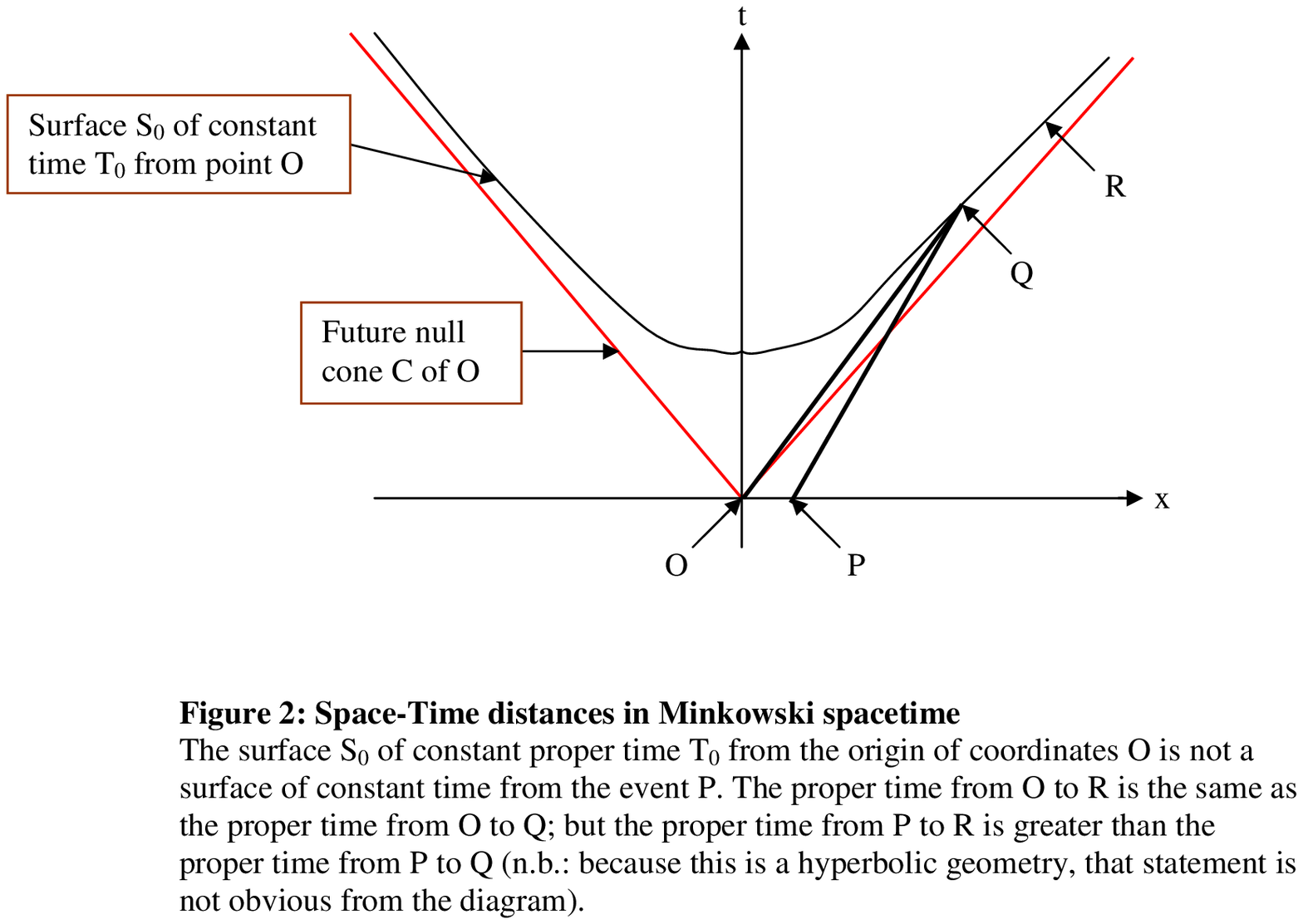}
\end{figure}
To show this in detail: let O be at $(T_O, X_O)=(0,0)$ and P be at
$(T_P, X_P)= (0,\epsilon)$ with $\epsilon > 0$. The surface $S_0$ is
given by
$$-T^2 + X^2 = -S_0^2. \eqno (4)$$
Thus Q is at $(T_Q,X_Q)$ where $X_Q=+\sqrt{T_Q^2 - S_0^2}$.
The space-time distance from P to Q is given by
$$-\tau_{PQ}^2 = -(T_P-T_Q)^2 + (X_P-X_Q)^2. \eqno (5)$$
Substituting for $(T_P, X_P,X_Q)$, (5) becomes
$$\tau_{PQ}^2 = \left(S_0^2-\epsilon^2\right)+2\epsilon \sqrt{T_Q^2 -S_0^2}.$$
Applying the same calculation to R and subtracting,
$$\tau_{PR}^2 -\tau_{PQ}^2 = 2\epsilon \left(\sqrt{T_R^2 -S_0^2}-\sqrt{T_Q^2 -S_0^2}\right), \eqno (6)$$
which diverges as $T_R \rightarrow \infty$ for fixed $P$ and $Q$,
provided $\epsilon \neq 0$ (i.e. P is not at O); for large $T_R$,
$$\tau_{PR}^2 \simeq 2\epsilon T_R. \eqno (7)$$
Thus a process that takes a minimum proper time $\tau_1$ to occur,
but depending on all the points in an initial domain $U = (0,x)$, $0
< x < \epsilon$, in the surface $T = 0$ will gradually progress up
the points in the surface $S_0 = const$ and will take an arbitrarily
long time to reach the furthest points in this surface, no matter
how small $\epsilon$ is, provided it is non-zero. It will take an
infinite time to affect all points in the surface in the true
meaning of the word infinity: it will always be progressing to
further and further points, and will never complete the process in
any finite time. If however $\epsilon = 0$ (i.e. P is at O), the
originating events are of zero spacelike extent and we recover
$\tau_{PR}^2 = \tau_{PQ}^2$ (both Q and R are at the same proper
time from O); all events on the surface occur simultaneously in
terms of proper time along the world lines from O. But this exactly
pointlike origin of later physical events is unphysical; it cannot
occur in practice.

\subsection{Bifurcate horizons}
The more general context context worth considering is that of
bifurcate Killing vectors, considered so clearly by Boyer
\cite{boy69}, which includes the previous example and the de
Sitter case considered in \cite{vilwin97} as special
cases\footnote{See for example Rindler \cite{rin01}, pages 49-54
and 306-309 respectively}.
The group of isometries acts as an isotropy group about the points
in the bifurcate Killing horizon, which are fixed points of the
group. As the group acts on the points in the space-time, the proper
time $\tau(O,P_i)$ from the origin $O$ to images $P_i$ of a point in
the surface of constant proper time from the origin stays invariant
(Figure 3). Similarly the proper time $\tau(Q_i,P_i)$ from images
$Q_i$ at constant spatial distance from the origin of a point $Q_1$,
to images $P_i$ of a point $P_1$ in the surface of constant time
from O, stays invariant (Figure 4). However the proper time
$\tau(Q_1,Q_i)$ from $Q_1$ to $Q_i$ increases linearly with the
group parameter $\xi$ along the timelike group orbits in the
surfaces of constant distance from the origin. The proper time
$\tau(P_1,Q_i)$ is greater than the sum of the times $\tau(P_1,P_2)$
and $\tau(P_2,Q_2)$ (as a timelike geodesic without conjugate points
is longer than any other curve between its endpoints). Hence
$$ \tau(P_1,Q_2) > \tau(P_1,P_2) + \tau(P_2,Q_2) = \alpha \xi + \tau(P_1,Q_1) \eqno (8)$$
for some constant $\alpha$; thus $\tau(P_1,Q_2) \rightarrow
\infty$ as $\xi \rightarrow \infty$ and the proper time from $P_1$
to points on the surface $T=0$ is unbounded in concordance with
(7). The physical conclusions are the same as in the previous
case. In particular this argument applies to the De Sitter
Universe, where the static frame represents the transition to
spatial surfaces of constant negative curvature.

\begin{figure}
\includegraphics[width=15cm]{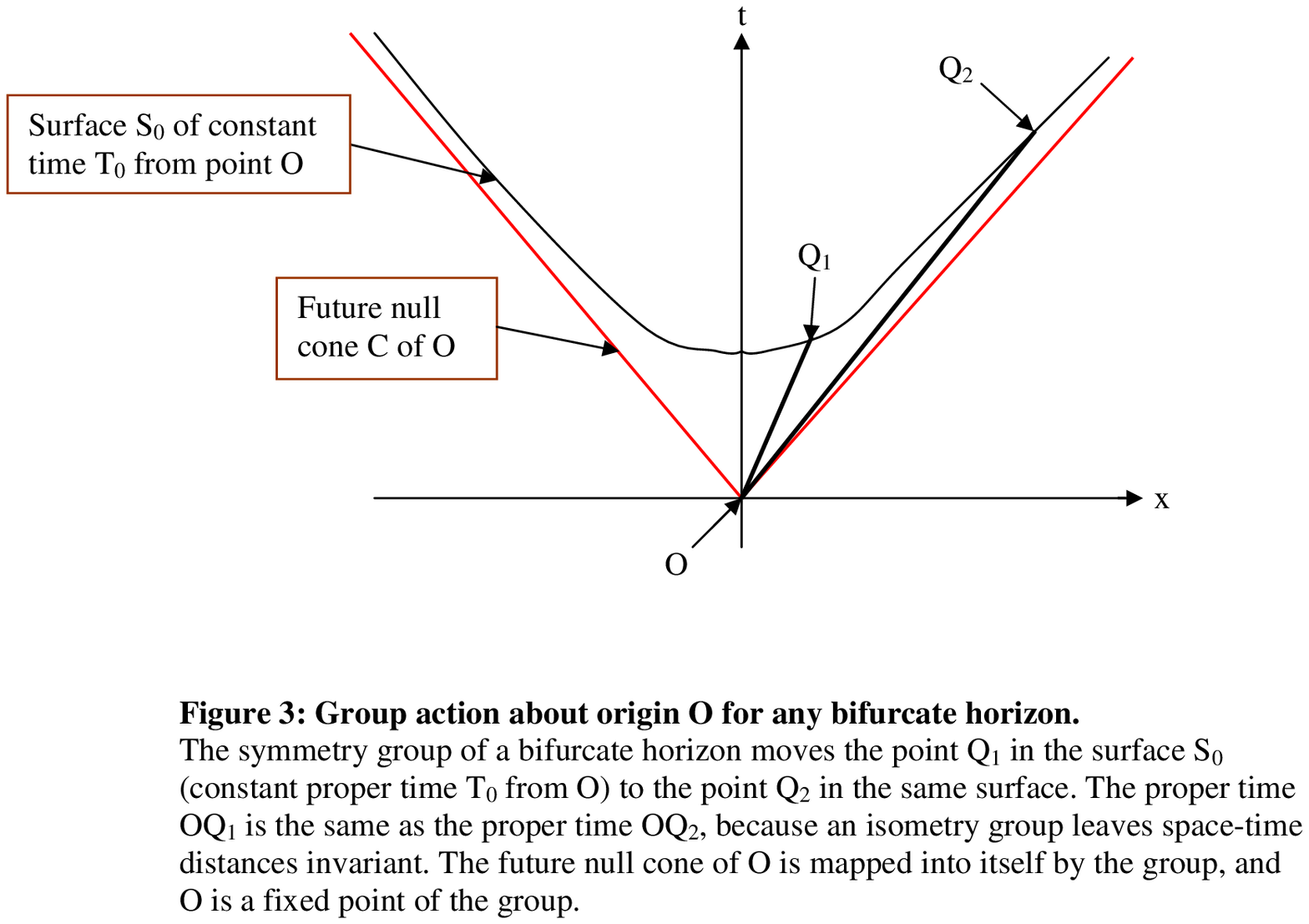}
\end{figure}

\begin{figure}
\includegraphics[width=15cm]{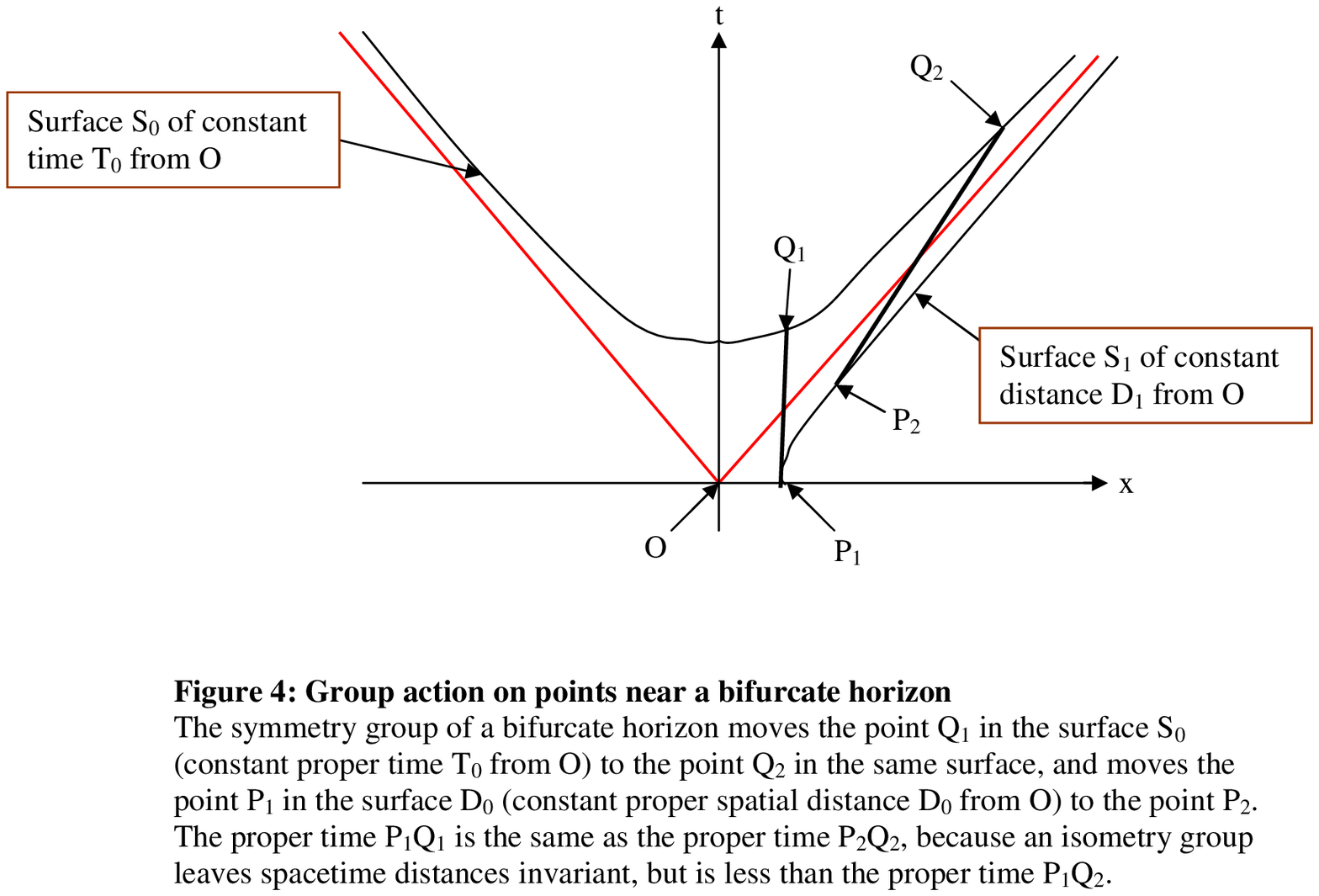}
\end{figure}

\begin{figure}
\includegraphics[width=15cm]{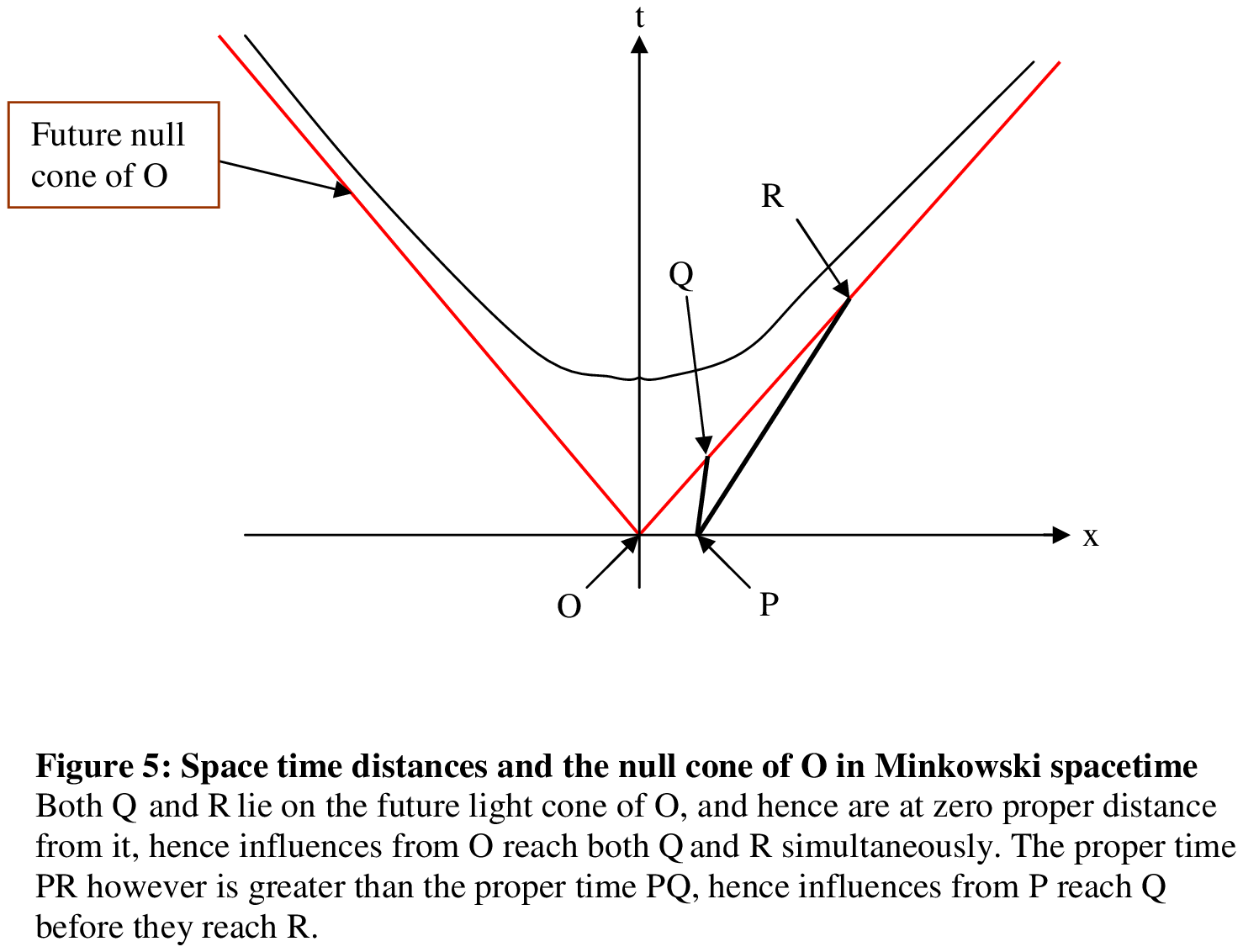}
\end{figure}

\subsection{The null cone revisited}
The same argument applies to the null cone, see Figure 5 for the
Minkowski spacetime situation. The set of points represented by the
null cone is instantaneous for the origin O, but more distant points
on the null cone are at larger and larger proper distances from any
point $P= (0,\epsilon)$ in the surface $T=0$, no matter how small
$\epsilon$, provided $\epsilon \neq 0$. Essentially the same
argument applies in both Figure 2 and Figure 5; hence events unroll
along the null cone too, because they are based in what occurs in a
finite rather than a pointlike domain in the initial surface $T=0$. In
fact the same Equations (4) - (7) apply as before, but now with $S_0
= 0$. Thus we get

$$\tau_{PR}^2 = \tau_{PQ}^2 + 2\epsilon \left(T_R - T_Q \right), \eqno (9)$$
again leading to (7) for large $T_R$.\\

This argument will again extend to the case of any bifurcate Killing
horizon as defined in \cite{boy69}. Figure 6 is the relevant
diagram, with the same argument as above in relation to Figure 4
leading to the same result as before, but now as regards points on
the light cone.

\begin{figure}
\includegraphics[width=15cm]{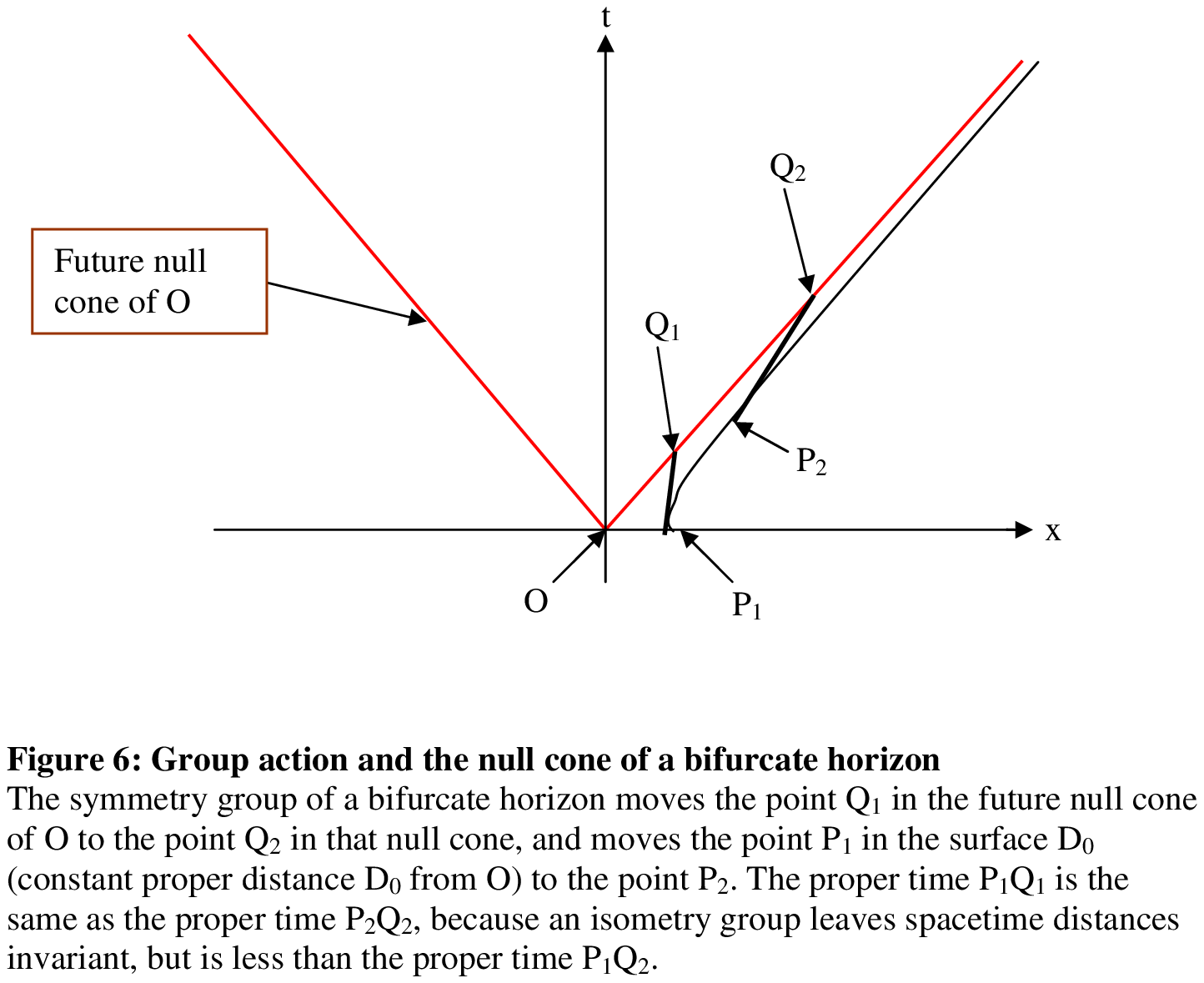}
\end{figure}

\section{Conclusion}
An infinite set of universes never exists at any finite time in
chaotic inflation - it is a state that is never attained, but is
what the physical situation tends to as time progresses. Similarly
spatial infinities cannot occur at any finite time in any one of the
universe domains resulting from quantum tunneling; rather they may
be the state that the models tend to as time goes to infinity. This
is so even though the hyperbolic infinite space sections are at unit
proper time from the origin, because what happens at later times is
not determined only by what happens at a spacetime point such as the
origin.\\

 Genuine physical processes will originate from domains
 with non-zero spatial extent. And as soon as this domain has finite extent,
 arbitrarily near points to the origin in the Minkowski surfaces of constant time
 $T = const$ are infinitely distant in proper time from the infinitely distant spatial events
 in the surfaces of constant proper time from the origin $S = const$, and it will take
 an infinite
 time for their consequences to develop to the farthest reaches of these spacelike surfaces.
 The putative spatial infinities will not come into being at any finite time. The key
 point is that the physics does not take place instantaneously in the
 spatial sections at constant proper distance from the origin, because it
 cannot have an \emph{exactly}
 pointlike physical origin. Consequently, the potential infinity is never actually
 attained in physical terms.
 It is always in the future of what actually exists.\\

 One of the main reasons that we study cosmology is because of our
 fascination with its philosophical aspects. The idea of
 infinite spatial sections has bizarre implications; if true, it plausibly implies that
 countless identical civilizations to ours are scattered in the infinite
 expanse of the cosmos, with semi-identical histories to ours replicated an
 infinite number of times out there \cite{ellbru79}. Vilenkin \emph{et al} claim
 this is a necessary outcome of current inflationary theories \cite{knoetal03,vil06}.
 We claim that this is
not the case, the catch lying in the idea that this is currently how
things are: that it is the state at the present instant. As
explained above, the real situation is that physical processes may
be such that eventually an infinite number of galaxies, stars,
planets, and civilizations will tend to come into existence; but
that state is not achieved at any finite time through the supposed
physical processes.\\

Any claims of actual existence of physical infinities in the real
universe should be treated with great caution (c.f. \cite{ell06a},
section 9.3.2), as emphasized by David Hilbert long ago
(\cite{hil64}, p. 151):

\begin{quotation}
\textit{``Our principal result is that the infinite is nowhere to
be found in reality. It neither exists in nature nor provides a
legitimate basis for rational thought . . . The role that remains
for the infinite to play is solely that of an idea . . . which
transcends all experience and which completes the concrete as a
totality . . ."}
\end{quotation}

\noindent Our results concur with this judgement.

\end{document}